\documentclass[english,floatfix,reprint,twocolumn,showpacs,aps, prl, superscriptaddress]{revtex4-2}

\usepackage{amsmath, physics}

\usepackage{hyperref}
\hypersetup{colorlinks=true, citecolor=blue, urlcolor=blue, linkcolor=blue, breaklinks=true}
\usepackage{amssymb}
\usepackage{graphicx, xcolor}
\usepackage{siunitx}
\usepackage{booktabs, footnote}

\newcommand{\dg}{^{\dagger }}

\newcommand{\vect}[1]{\boldsymbol{\mathbf{#1}}}

\begin{document}
\title{Distinct effect of Kondo physics on crystal field splitting in electron and spin spectroscopies}

\author{M. Kornja\v ca}
\affiliation{Ames Laboratory, U.S.~Department of Energy, Ames, Iowa 50011}
\affiliation{Department of Physics and Astronomy, Iowa State University, Ames, IA, 50011}

\author{R. Flint}
\affiliation{Ames Laboratory, U.S.~Department of Energy, Ames, Iowa 50011}
\affiliation{Department of Physics and Astronomy, Iowa State University, Ames, IA, 50011}

\date{\today} 

\begin{abstract}


Magnetic anisotropy is a key feature of rare earth materials from permanent magnets to heavy fermions.  We explore the complex interplay of Kondo physics and anisotropy, and their effect on different experimental probes of magnetic anisotropy in a minimal $J = 3/2$ Anderson impurity model using numerical renormalization group.  While anisotropy suppresses Kondo screening, virtual valence fluctuations enhance the anisotropy. We find distinct renormalization of the magnetic anisotropy measured via dynamical spin response (inelastic neutron scattering) versus electronic excitations in the impurity spectral function (resonant inelastic x-rays and scanning tunneling spectroscopy).  The two measurement types have different responses and dependences on the temperature and Kondo scales.


\end{abstract}

\keywords{$4f$}
\maketitle

Magnetic anisotropy in rare earth intermetallics can be exceptionally strong and useful, particularly for high performance ferromagnets\cite{McCallum2014}, where it allows high coercivities.  It is also a key tuning parameter for topological materials \cite{Lee2023}, and heavy fermion magnetism, superconductivity \cite{Pagliuso2002} and quantum criticality \cite{Chen2017,Kirchner2020}.  Modeling and measuring this anisotropy is an important, challenging problem, particularly where Kondo screening renormalizes the anisotropy, as in hard axis ferromagnetic ordering \cite{Hafner2019}.  While the effect of anisotropy on Kondo screening has been explored \cite{Hanzawa1985,Bickers1987,Anders2006}, the effect of Kondo screening on the anisotropy, particularly spectroscopically, is still an open question.  Here, we revisit the Kondo renormalization of magnetic anisotropy as measured by magnetic and electronic probes, and show that different experimental techniques measure distinct renormalized quantities, which allows more insight into the underlying physics. 

Localized rare earth $f$ electrons have large Coulomb interactions ($\sim$ eV) and spin-orbit couplings ($\sim$ 100 meV) that lead to well-defined total $J$ multiplets that can undergo $SU(N = 2J+1)$ Kondo screening or be split by the weak crystalline electronic field (CEF) ($\sim$ meV) into doublets with highly anisotropic moments that may be screened at lower temperatures. The CEF splitting protects this magnetic anisotropy, and can be measured by fitting thermodynamic data or by inelastic neutron scattering (INS) \cite{Willers2010}, photoemission \cite{Allen2005,Duan2019,Chen2018,Patil2016,Vyalikh2010,Reinert2001,Ehm2007}, resonant inelastic x-ray scattering (RIXS) \cite{Amorese2016,Amorese2018a,Amorese2018b,Amorese2019,Chiogo2022,Amorese2023,Christovam2024}, time-domain THz spectroscopy \cite{Pal2019,Shee2024} and scanning tunneling microscopy (STM) \cite{Jiao2016,Ernst2011,Haze2019}.  Even without Kondo physics, computing the CEF splitting in intermetallics is difficult \cite{zhou2009prb,Lee2024}, making accurate measurements very important. However, Kondo broadening of the spectral peaks \cite{Christianson2004} makes these measurements difficult, and sometimes makes it hard to reconcile results from different techniques \cite{Amorese2016}. 

The effect of CEF splitting, $\Delta$ on impurity $SU(N)$ Kondo physics has been theoretically studied with a variety of analytical and numerical methods\cite{Schlottmann1984,Desgranges2015, Hanzawa1985,Bickers1987,Zevin1988,Han1997,Riseborough2003,Kroha2003,Moreno2005,Anders2006}. The simplest $SU(4)$ case captures the key physics.  Without CEF splitting, there is a large $SU(4)$ Kondo temperature, $T_K$ where the entire quartet is screened. The $SU(4)$ fixed point is marginally unstable to any $\Delta$, which suppresses the Kondo screening until the excited doublet is unoccupied and the ground state doublet is screened at an exponentially suppressed $SU(2)$ $T_K$.  This crossover from $SU(4)$ to $SU(2)$ Kondo behavior occurs as a function of $\Delta/T_K$, and can be seen in thermodynamic quantities \cite{Schlottmann1989, Anders2006,Desgranges2015}.  Impurity results also capture the high temperature, incoherent physics of the Kondo lattice, where the Kondo resonances become coherent heavy fermion bands at low temperature.  The effect of Kondo physics on the magnetic anisotropy is less well studied, and also particularly relevant as several Kondo ferromagnets unexpectedly order along the nominally ``hard'' magnetic axis \cite{Hafner2019}.  There are likely significant differences in the role of anisotropy in the $SU(N)$ models relevant for rare earths compared to the $S \geq 1$ $SU(2)$ Kondo models relevant for transition metals like Co \cite{Otte2008,Zitko2009b} or organic molecules \cite{Kawai2012,Blesio2023}. For example, while theoretical studies of $S \geq 1$ SU(2) Kondo models with single ion anisotropy show that fluctuations can invert the anisotropy \cite{Kwasigroch2022}, we do not find CEF inversions with temperature in our $SU(4)$ model.

In this paper, we explore the effect of valence fluctuations on magnetic anisotropy using numerical renormalization group (NRG) \cite{Wilson1975,Bulla2008,Anders2006} studies of a minimal anisotropic $SU(4)$ Anderson impurity model with a CEF split $J = 3/2$ quartet, where we find two distinct spectroscopic signatures of the CEF splitting.  The first is a splitting of the Kondo resonance in the impurity spectral function, $A(\omega)$ corresponding to a renormalized CEF splitting ($\Delta^*$) between the excited, magnetic ``broken Kondo singlet'' states that are observable by effectively electronic probes like photoemission, RIXS and STM. The second is a peak in the INS response, $S(\omega)$ corresponding to excitations between the ground and excited state Kondo singlets ($\omega_\perp$).  The two features are related, but clearly distinct, as they are affected differently by temperature, valence fluctuations, and the ratio of $\Delta/T_K$.
The renormalized CEF scale, $\Delta^*$ is generically enhanced by the development of valence fluctuations at intermediate temperatures, while the INS peak, $\omega_\perp$ tends to be monotonically enhanced with increasing temperature near $T_K$.  Interestingly, the $\Delta^*$ enhancement mostly happens well above $T_K$ where the moments remain largely unscreened, suggesting that Kondo physics could enhance the anisotropy near the operating temperatures of permanent magnets.

\begin{figure}[!ht]
    \centering
    \includegraphics[width=0.48\textwidth]{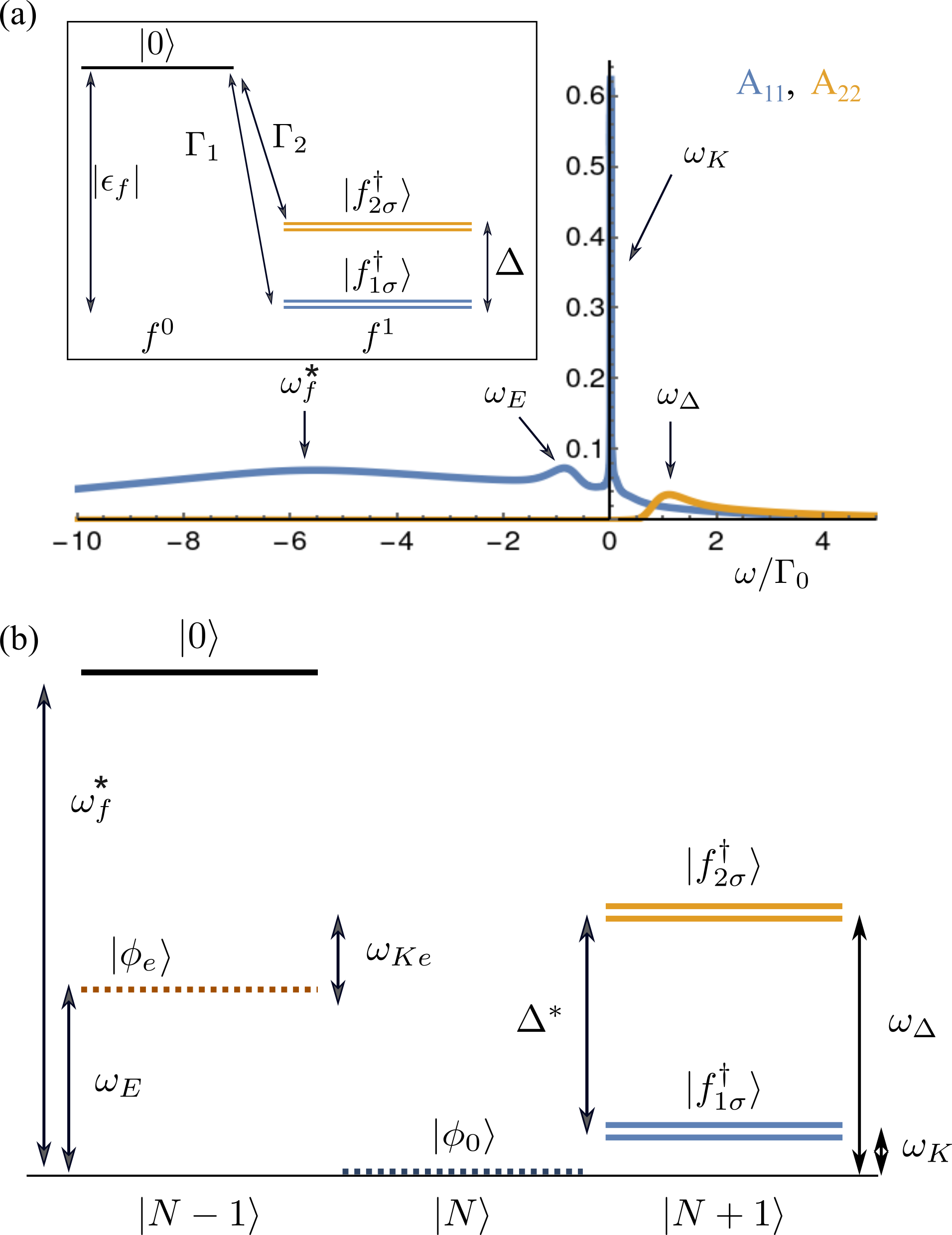}
    \caption{(a) Example impurity spectral function, with the ground ($A_{11}$) and excited state ($A_{22}$) spectral functions in blue and orange, for $|\epsilon_f|/\Gamma_0=8$, $\Delta/\Gamma_0=1$, $T/\Gamma_0 = 0.0001$, with $T=0$ empty state occupancy, $n_0 = .06$. (Inset) The $J = 3/2$ infinite-$U$  Anderson model has two $f^1$ doublets split by $\Delta$, $\ket{f\dg_{\alpha\sigma}} = \ket{J_z = \pm \frac{3}{2}}, \ket{J_z = \pm \frac{1}{2}}$, and an empty state ($f^0$) with energy cost $|\epsilon_f|$; other atomic configurations are forbidden. The doublets hybridize, $\Gamma_{1,2}$ through two conduction channels, allowing for perfect screening. (b) The four spectral function peaks capture charged excitations ($N\rightarrow N\pm1$) out of the ground state Kondo singlet, $|\phi_0\rangle$ \cite{Bickers1987}. The sharp central peak, $\omega_{K}$ is the Kondo resonance for breaking the Kondo singlet by adding an $f_1$; $\omega_\Delta = \omega_K + \Delta^*$ corresponds to adding an $f_2$; $\omega_{f0} = |\epsilon_f^*|$ is the empty state; and $\omega_{E} = \Delta^*-\omega_{EK}$ corresponds to removing an $f_1$ into an energy  overlapping the excited Kondo singlet, $|\phi_e\rangle$, with energy lowered from the uncorrelated excited doublet by $\omega_{Ke}$.} 
    \label{fig:spectral}
\end{figure}

We consider a minimal $J=3/2$ Anderson model with tetragonal or hexagonal CEFs splitting the quartet into doublets: $|\Gamma_7\pm\rangle = |J_z\!=\!\pm 3/2\rangle$ and $|\Gamma_6\pm \rangle = |J_z\! =\! \pm 1/2\rangle$:
\begin{align}\label{eq:AndHam}
    H &= \sum_{\vect{k}\alpha\sigma}\epsilon(\vect{k}) c_{\vect{k}\alpha\sigma}^\dagger c_{\vect{k}\alpha\sigma}+\epsilon_{f}n_{f1}+(\epsilon_{f}+\Delta)n_{f2}\cr
    &\!\!+U\!\!\!\!\sum_{\alpha\sigma \neq \alpha'\sigma'}\!\!\!\!n_{f\alpha\sigma}n_{f\alpha'\sigma'}
    +\!\sum_{\vect{k}\sigma\alpha}\!\left[V_{\alpha}(\vect{k})c_{\vect{k}\alpha\sigma}^\dagger f_{\alpha\sigma} + h.c.\right].
\end{align}
$f\dg_{\alpha\sigma}$ creates an impurity fermion in doublet $\alpha$ with spin $\sigma = \pm$; $n_{f\alpha\sigma} = f\dg_{\alpha\sigma} f_{\alpha\sigma}$ and $n_{f\alpha} = \sum_\sigma n_{f\alpha\sigma}$.  $c\dg_{\vect{k}\alpha\sigma}$ creates a conduction electron that hybridizes with doublet $\alpha$, with hybridization $V_{\alpha}$. Each doublet hybridizes through a single conduction channel, resulting in perfect screening.  $\epsilon_f < 0$ is the impurity chemical potential, and $\Delta$ the bare CEF splitting.  We include an impurity interaction, $U$ for clarity, but take $U$ to infinity, prohibiting double occupancy.  The relevant energies and parameters are sketched in the inset of Fig. \ref{fig:spectral}.  The impurity physics is completely described by the local Hamiltonian and hybridization functions \cite{Bulla2008}:
\begin{equation}\label{eq:hybrd}
   \Gamma_{\alpha}(\omega)=\pi\sum_{\vect{k}}\delta\left(\omega-\epsilon(\vect{k})\right) |V_{\alpha}(\vect{k})|^2.
\end{equation}

\begin{table}[bt]
  \caption{
    Relevant $N$ particle many body states and their energies.  There are: two magnetic doublets corresponding to the unscreened magnetic states, with renormalized splitting, $\Delta^* = \omega_\Delta - \omega_K$; two Kondo singlets, whose energies are lowered from their respective magnetic doublets by their Kondo energy, $\omega_K$, $\omega_{Ke}$; and a trivial 4f$^0$ singlet at $\omega_f^*$.
    The excited Kondo singlet is at $\omega_E = \Delta^* - \omega_{Ke}$. The labels ($\omega_K, \omega_E, \omega_\Delta$) are used for peaks measured in the impurity spectral function, while ($\omega_{11}, \omega_{\perp}, \omega_{22}$) label the corresponding excitations in the dynamic susceptibilities defined below eqn \ref{structure}.
  }
\bgroup \def\arraystretch{1.3}
\begin{tabular*}{\linewidth}{c l @{\extracolsep{\fill}}l}
\hline\hline
Energy $\quad$ & $\qquad$ State & Ket\\
\hline
0 & Ground state Kondo singlet & $|\phi_0\rangle$ \\
$\omega_K$ ($\omega_{11})$ & Unscreened ground state doublet & $|\mathrm{FS_{N-1}}; f_{1\sigma}\dg\rangle$ \\
$\omega_{E}$ ($\omega_{\perp})$ & Excited Kondo singlet & $|\phi_e\rangle$ \\
$\omega_{\Delta}$ ($\omega_{22})$ & Unscreened excited doublet  & $|\mathrm{FS_{N-1}}; f_{2\sigma}\dg\rangle$ \\
$\omega_f^*$ & Empty state singlet & $|\mathrm{FS_{N}}; f^0\rangle$\\
\hline\hline
\end{tabular*}
\egroup
\label{tbl:states}
\end{table}

To understand the Kondo physics, we first discuss the relevant many body states with $N$ electrons.  Without hybridization, there are two magnetic doublets, $|\mathrm{FS_{N-1}}; f_{\alpha\sigma}\dg\rangle$ separated by $\Delta$, and the empty $f$ singlet, $|\mathrm{FS_{N}}; f^0\rangle$.  With weak hybridization, each magnetic doublet admixes in some of the empty state to form a Kondo singlet state, which we call $|\phi_0\rangle$ and $|\phi_e\rangle$, respectively.  With larger hybridization, some of the other magnetic doublet also gets admixed in to the Kondo singlet.  The ground state Kondo singlet is,
\begin{equation}
    |\phi_0\rangle = \left[u +\!\! \sum_{\vect{k}\alpha\sigma}v_{\vect{k}\alpha}f\dg_{\alpha\sigma}c_{\vect{k}\alpha\sigma} \right]|\mathrm{FS}_N; f^0\rangle,
\end{equation}
where $|\mathrm{FS}_N\rangle$ is the $N$ conduction electron Fermi sea.  The Kondo singlet is a superposition of three components: the empty $f$ state, $|f^0\rangle$ and $N$ conduction electrons (weight $n_0 = |u|^2$);  one $f$-electron in orbital $\alpha = 1$, $|f_{1\sigma}\dg\rangle$ and $N-1$ conduction electrons (weight $n_1 = \sum_{\vect{k}} |v_{\vect{k}1}|^2$); and one $f$-electron in orbital $\alpha = 2$, $|f_{2\sigma}\dg\rangle$ and $N-1$ conduction electrons (weight $n_2 = \sum_{\vect{k}} |v_{\vect{k}2}|^2$).  The system gains energy $\omega_K$ by forming this Kondo singlet, compared to the unscreened magnetic $|f_{1\sigma}\dg\rangle$ state with $N-1$ conduction electrons.

The excited state Kondo singlet has a similar form,
\begin{equation}
    |\phi_e\rangle = \left[u_e +\!\! \sum_{\vect{k}\alpha\sigma}v_{e,\vect{k}\alpha}f\dg_{\alpha\sigma}c_{\vect{k}\alpha\sigma} \right]|\mathrm{FS}_N; f^0\rangle.
\end{equation}
This excited many body state will be at $\omega_E = \Delta^* - \omega_{Ke}$, with energy lowered from the excited CEF level by the excited state Kondo scale, $\omega_{Ke}$,
While we label these ground and excited Kondo singlets, they do not generally have distinct orbital characters; this labeling is only valid for $\Delta \gtrsim \omega_K$.   For small $\Delta$, there is only one Kondo singlet that splits with increasing $\Delta$. 

\begin{figure}[!ht]
    \centering
    \includegraphics[width=.48\textwidth]{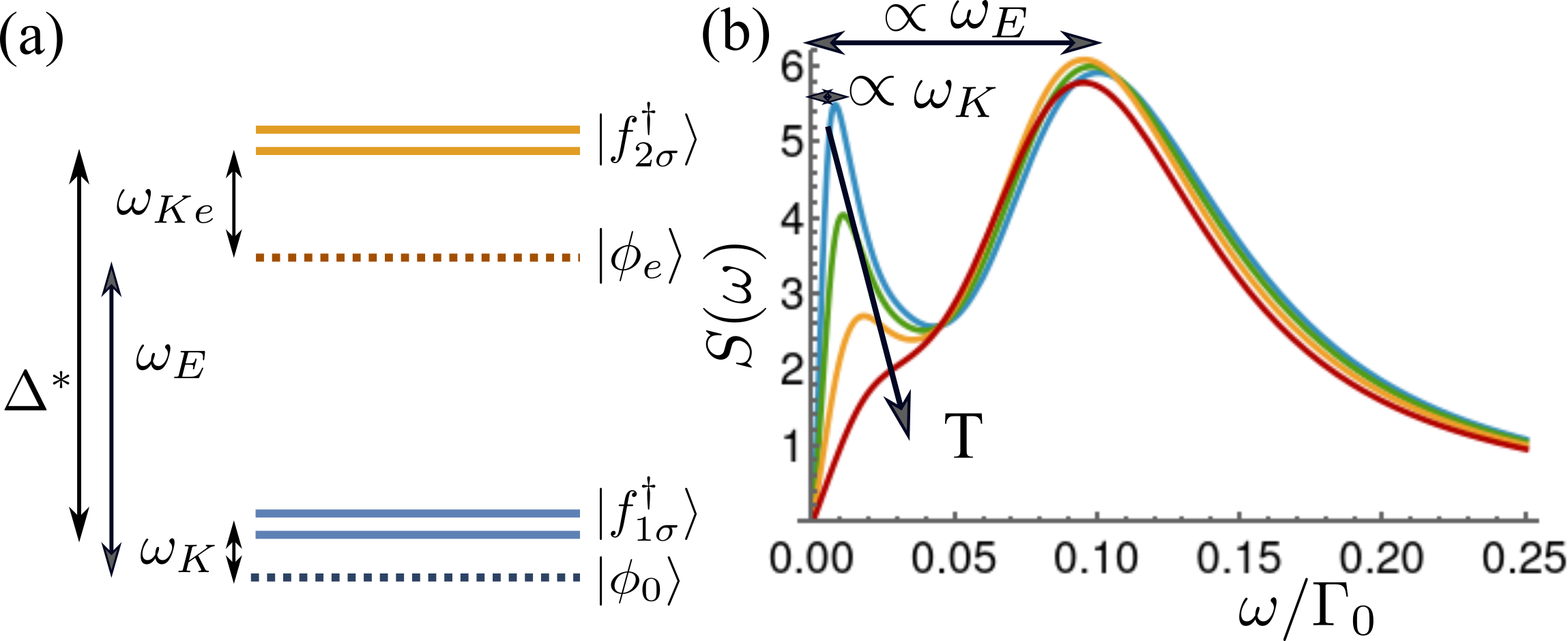}
    \caption{(a) Relevant energy levels for neutral magnetic excitations out of the ground state Kondo singlet, $|\phi_0\rangle$.  Low-lying magnetic excitations to broken Kondo singlet states, $|f_{\alpha\sigma}\dg\rangle$ manifest as peaks in $\chi_{\alpha\alpha}''(\omega)$, while  $\chi_\perp''(\omega)$ appears to be dominated by excitations between the ground and excited state Kondo singlets. (b) Temperature dependence of the INS structure factor, $S(\omega)$ for $\Delta/\Gamma_0 = 0.1$, $|\epsilon_f|/\Gamma_0 = 8$.  Temperatures are $0.003\Gamma_0$ (blue), $0.006\Gamma_0$ (green), $0.01\Gamma_0$ (orange), $0.02\Gamma_0$ (red), all at or above $T_K \sim 0.003\Gamma_0$, defined as where the empty state occupancy, $n_0(T)$ saturates ($n_0 = .08$).  The lower peak is associated with the Kondo resonance ($\propto \omega_K$), and shifts and broadens significantly with $T$, while the upper CEF peak ($\propto \omega_E$) shifts slightly without much broadening.}
    \label{fig:INS}
\end{figure}

We solve this model using NRG \cite{Wilson1975, Krishna1980, Hofstetter2000, Bulla2008}, employing the NRG Ljubljana open source package \cite{Zitko2009a, Zitko2021}. In NRG, the Anderson impurity model is mapped to a finite chain with a logarithmic discretization of the continuous conduction bath, with discretization parameter $\Lambda > 1$. The exact limit corresponds to $\Lambda \rightarrow 1$, but larger $\Lambda$ can allow for more efficient calculations if the resulting finite size effects are minimized by averaging over $N_z$ discretization realizations; we use the  Campo-Olivera scheme as implemented by NRG Ljubljana \cite{Campo2005, Zitko2009a}. The diagonalization of the chain Hamiltonian is performed iteratively, with a basis truncation at each iteration that keeps maximum $m$ states.  NRG Ljubljana enforces the relevant symmetries, which lowers the required $m$; as we do not consider external fields, we use the full $SU(2)$ spin symmetries for each doublet, and the overall $U(1)$ charge symmetry.  Unless otherwise indicated, we set $\Lambda = 4$, $N_z = 4$ and $m = 1500$, which allows for good convergence of both thermodynamic quantities and spectroscopic functions.  We choose a flat conduction electron density of states, and isotropic, $V_1 = V_2 = V$ hybridization functions constant within $[-D,D]$, with $D$ the conduction bandwidth. We also fix $U = 100$, $D=1$, $V = 0.113$, with $\Gamma_0 = \frac{1}{2D}\pi V^2 = 0.02$, and vary $|\epsilon_f|$ and $\Delta$.  Our results are consistent with previous work on the same model \cite{Anders2006}, where applicable. Other checks and error estimates are discussed in the Supplementary Material (SM) \cite{SM}.  The many body states introduced above can be probed by both electronic and spin spectroscopies, which are captured by peaks in the frequency dependent impurity spectral functions (Fig. \ref{fig:spectral}) and dynamical spin susceptibilities (Fig. \ref{fig:INS}).

\begin{figure}[!ht]
    \centering
    \includegraphics[width=.5\textwidth]{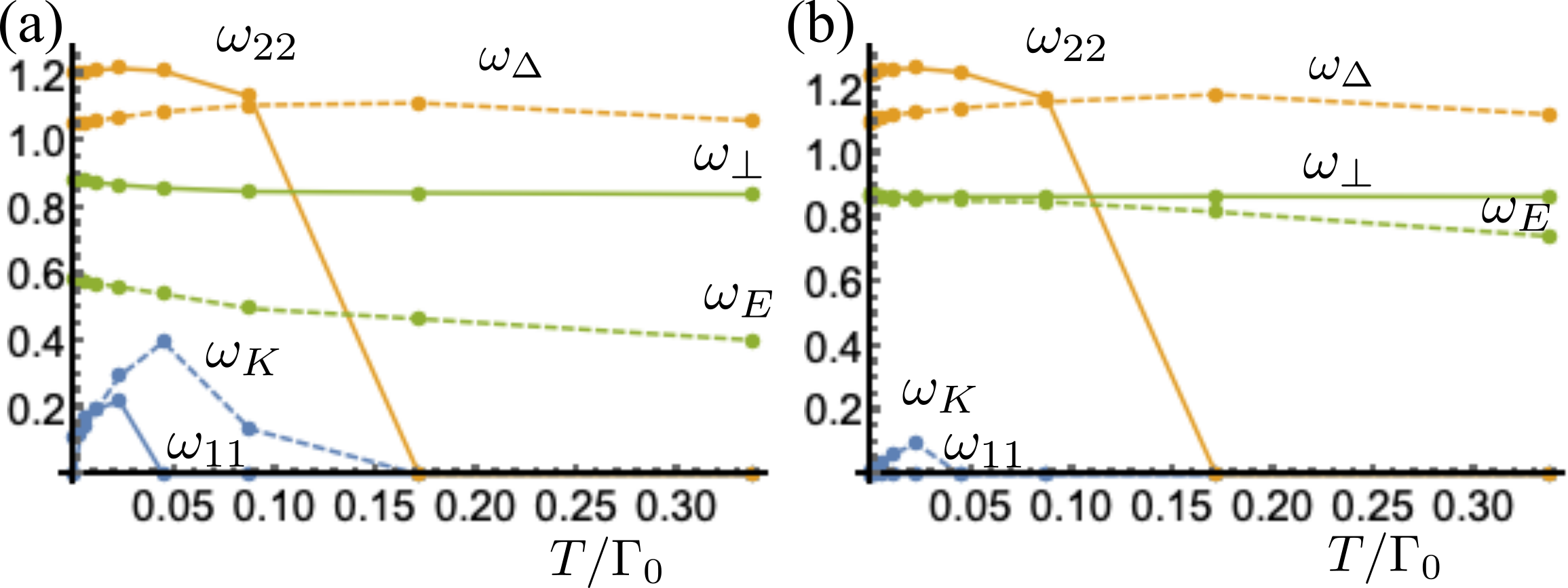}
    \caption{Temperature dependence of the peaks in $A(\omega)$ [$\omega_K$, $\omega_\Delta$, $\omega_E$] and $\chi''(\omega)/\omega$ [$\omega_{11}$, $\omega_{22}$, $\omega_\perp$], for (a) $\Delta/\Gamma_0 = 1$, $\epsilon_f /\Gamma_0= 6$, $n_0 = .13$; (b) $\Delta/\Gamma_0 = 1$, $|\epsilon_f|/\Gamma_0 = 8$, $n_0 = .06$. 
    The peaks in $\chi''(\omega)/\omega$ ($\chi$ = $\chi_{\alpha\alpha}$, $\chi_\perp$) are non-zero when inelastic and zero when quasi-elastic. When inelastic, the spectral and spin susceptibility peaks are proportional; the Kondo resonances, $\omega_K$ and $\omega_{11}$ (blue) are $\times 20$ for clarity.  The monotonic in $T$ decrease of $\omega_\perp$ and $\omega_E$ (green) is similar, and distinct from the other two more dome-like sets, but $\omega_\perp/\omega_E$ varies markedly with $\epsilon_f$, likely due to increasing $n_0$.}
    \label{fig:peaks}
\end{figure}

First, we consider the impurity spectral functions for both ground and excited state electrons, $f_1$ and $f_2$.  At $T = 0$ these are the probability to add or remove electrons in orbital $\alpha$ from the ground state Kondo singlet, $|\phi_0\rangle$ into many body states $|\lambda\rangle$,
\begin{equation}
    A_{\alpha \alpha}(\omega) = \sum_\sigma \sum_\lambda |\langle \lambda|f\dg_{\alpha\sigma}|\phi_0\rangle|^2 \delta(\omega+E_0-E_\lambda).
\end{equation}
There are up to four distinct peaks in the spectral functions \cite{Bickers1987}, broadly corresponding to the four other many body states as seen in Table \ref{tbl:states} and Fig. \ref{fig:spectral}.  There are two ways to add an electron to the ground state Kondo singlet ($\omega > 0$), thereby breaking it to obtain a magnetic doublet, $|\mathrm{FS}_{N}; f_{\alpha\sigma}\dg\rangle$; these are Kondo resonances at $\omega_K$ ($A_{11}$) or $\omega_\Delta$ ($A_{22}$).  There are two ways to remove an electron ($\omega < 0$): either into the empty state, $|\mathrm{FS}_{N-1};f^0\rangle$ at $\omega_{f}^*$ or into the empty part of the excited Kondo singlet with $N-1$ electrons at $\omega_E$; these processes contribute to both $A_{11}$ and $A_{22}$.  The temperature dependence of the peaks is shown in Fig. \ref{fig:peaks}.

The renormalized CEF splitting shows up as a splitting of the two $\omega > 0$ peaks that add either $f_1$ or $f_2$ electrons to $|\phi_0\rangle$, with $f\dg_{\alpha \sigma}|\phi_0\rangle \propto |\mathrm{FS_{N}};f\dg_{\alpha \sigma}\rangle$. For $\Delta = 0$, there is a single Kondo resonance peak corresponding to the energy cost for breaking the ground state Kondo singlet, which is only possible because of the admixture of the empty state into the Kondo singlet as double occupancy is prohibited.   Finite $\Delta$ splits it into two peaks (at $\omega_K$ in $A_{11}$ and $\omega_\Delta = \Delta^* + \omega_K$ in $A_{22}$), similar to the effect of magnetic field for $SU(2)$ Kondo \cite{Rosch2003,Zitko2009b,Zitko2011b}.  Note that the occupancy of orbital $\alpha$ is $n_\alpha = \int_{-D}^0 A_{\alpha \alpha}(\omega)$, with $n_0 = 1- n_1 - n_2$.  The weight of each peak above $\omega = 0$ is $2 n_0$, where $n_0$ is the occupancy of the empty state, as there is $n_0$ ``room'' for additional $f$ electrons and four ($\alpha \sigma$) total ways to add those electrons.  At finite temperatures, electrons are added and removed from a thermal superposition of low energy states - particularly, the lower magnetic doublet starts to populate and the two Kondo resonance peaks will include excitations between magnetic states that tend to broaden and raise the peaks.  For large $\Delta$, the upper Kondo resonance peak coexists with many simple CEF excitations between the two doublets and is simply the non-interacting CEF peak.  The effective CEF splitting, $\Delta^*(T) = \omega_\Delta - \omega_K$ represents the difference in how the peaks smear, making it a good proxy for how the low-lying \emph{magnetic} states split with increasing $T$. 

The asymmetric lower Kondo resonance is phenomenologically fit by a generalized Fano-Frota form \cite{Zitko2011b,Pruser2012},
\begin{equation}
    A_{GF}(\omega) = A_0 \mathrm{Re}\left\{\mathrm{e}^{i\phi} \left[\frac{i \Gamma}{\omega-\omega_0 +i\Gamma}\right]^{a}\right\}.
\end{equation}
The Frota form, $\frac{2}{\pi\Gamma_0}\mathrm{Re}\left[\sqrt{\frac{i \Gamma}{\omega-\omega_0 +i \Gamma}}\right]$ has long been used to model Kondo resonance peaks in the symmetric $SU(2)$ Anderson model, with the square-root motivated by orthogonality catastrophe physics \cite{Frota1992,Doniach1970}.  We generalize the power law $a = 1-2(\delta/\pi)^2$, $\delta = \pi/N$ from $N = 2$ to interpolate between $SU(2)$ and $SU(4)$ with a more general power $1/2 < a < 7/8$.  $\phi$ allows a Fano asymmetry, which was introduced to fit STM Kondo resonances, where the asymmetry comes from co-tunneling \cite{Zitko2011b,Pruser2012}; here the asymmetry comes from the infinite $U$ \cite{Zhuravlev2004}.  We fit this peak with $A_{GF}(\omega)$ and define $\omega_K$, $\Gamma_K$ as the peak and half-width-half-maximum of the fitted curve.  
$\omega_K(T)$ initially increases with $T$ and then decreases as the peak broadens, persisting to $\sim 10 T_K$.  

\begin{figure}[!ht]
    \centering
    \includegraphics[width=.48\textwidth]{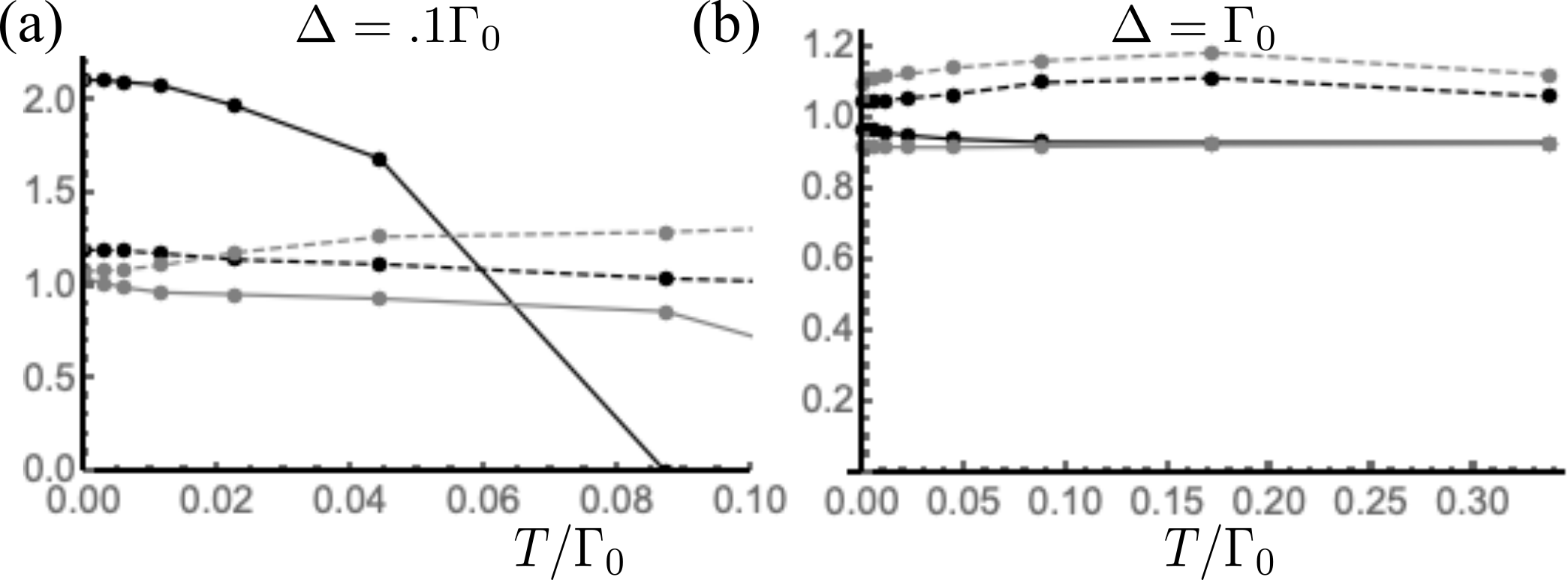}
    \caption{Temperature dependence of two measures of CEF splitting: $\Delta^*(T)$ (dashed lines) and $\omega_\perp$ (solid lines), both in units of the bare $\Delta$. Colors indicate different $|\epsilon_f|/\Gamma_0 = 6$ (black), $8$ (gray).  (a) shows $\Delta/\Gamma_0 = 0.1$, where the Kondo scales are $\omega_K/\Gamma_0 = 0.08, 0.002$.  The temperature scale cuts off at $T = \Delta$, where the peaks are very broad and $\chi_\perp$ becomes quasi-elastic.  $|\epsilon_f|/\Gamma_0 = 6$ exemplifies the behavior for $\Delta \lesssim \omega_K$, where $\omega_\perp$ is substantially enhanced by valence fluctuations ($n_0 = .17$), particularly compared to $\Delta^*$.
    (b) shows $\Delta/\Gamma_0 = 1$, where $\omega_K/\Gamma_0 = 0.005, 0.0001$.  Higher $T_K$'s lead to larger enhancement of $\omega_\perp$ at low $T$, agreeing with scaling predictions, while lower $T_K$'s lead to more $\Delta^*$ enhancement.  
   }
    \label{fig:enhance}
\end{figure}

The upper Kondo resonance or CEF peak is located at $\omega_\Delta \geq \omega_K$, with width $\Gamma_\Delta > \Gamma_K$, where $\Gamma_\Delta$ is renormalized by interactions to be much narrower than the non-interacting width, $\Gamma_0$.
We interpret the splitting, $\Delta^* = \omega_\Delta - \omega_K$ as the renormalized CEF scale, which is enhanced by up to 40\%  from the bare $\Delta$, as seen in Fig. \ref{fig:enhance}.  For $\Delta \lesssim \omega_K$, the peak is well-fit by the generalized Fano-Frota form, but not for larger $\Delta$, where it is not really a Kondo resonance \cite{SM}.  $\omega_\Delta$ also increases with $T$, and the peak broadens and persists well above $T \sim \Delta, T_K$.

Now we turn to $\omega < 0$.  Both $f_1$ and $f_2$ electrons can be removed from the Kondo singlet into empty ($f^0$) states.  The first peak is associated with the usual broad $f^0$ state, whose energy is renormalized to $\omega_{f0} = -\epsilon_f^*$. 
The second peak is associated with the excited Kondo singlet at $\omega = -\omega_E = -(\Delta^*-\omega_{Ke})$, corresponding to excitations into the $f^0$ ($u_e$) part of the unoccupied excited Kondo singlet, which gives a pileup of $f^0$ weight $\omega_{Ke}$ below the renormalized $\Delta^*$ \cite{Bickers1987}. 
This peak is often discussed as a CEF ``side-band'' with $\omega_E = \omega_\Delta$ mirrored from $\omega > 0$ \cite{Reinert2001,Kroha2003,Ehm2007,Pal2019,Tosi2013}; while this is roughly true for large $\Delta$, generically $\omega_E$ is substantially smaller than $\omega_\Delta$, as $\omega_{Ke} \gg \omega_K$, and the two peaks have opposite $T$ dependences, with this peak narrowing slightly and moving towards $\omega = 0$ with increasing $T$.  

While the spectral function tells us about electronic excitations, the dynamical spin susceptibilities describe neutral magnetic excitations, primarily through the INS structure factor shown in Fig. \ref{fig:INS}: 
\begin{align}
    S(\omega) & \propto n_B(\omega)\left\{ \frac{1}{3}\left[\left(\frac{3}{2}\right)^2\chi''_{11}(\omega)+\left(\frac{1}{2}\right)^2\chi''_{22}(\omega)\right]\right.\cr
    & \left. +\frac{2}{3}\left[\chi''_{22}(\omega) + \left(\frac{\sqrt{3}}{2}\right)^2\chi''_\perp(\omega) \right]\right\}.
    \label{structure}
\end{align}
$S(\omega)$ is the powder-averaged neutron structure factor for $J=3/2$ in tetragonal or hexagonal symmetry, where the first, second terms are the $J_z$ and $J_x$ responses, respectively, and $n_B(\omega)$ is the Bose factor \cite{SM}.  Here, we fix the ground state doublet to be $|\pm 3/2\rangle$, giving Ising anisotropy. There are three component susceptibilities: two intra-doublet \cite{Anders2006},
$\chi_{\alpha\alpha}(\omega) = \int dt \mathrm{e}^{i \omega t} \langle S_{z,\alpha}(t)S_{z,\alpha}(0)\rangle$,
where $S_{z,\alpha} = \sum_{\sigma} \sigma f\dg_{\alpha\sigma} f_{\alpha\sigma}$, and the inter-doublet susceptibility, 
\begin{equation}
    \chi_\perp(\omega) = \int dt \langle J_\perp(t)J_\perp(0)\rangle \mathrm{e}^{-i \omega t}.
\end{equation}
$J_\perp = \sum_{\sigma} f\dg_{1\sigma}f_{2\sigma} + f\dg_{2\sigma}f_{1\sigma}$ allows transitions between the two doublets.  In the absence of magnetic field, $\chi_{\alpha\alpha}$ are $SU(2)$ symmetric, and $\chi_{12}$ is zero (as is $A_{12}$). $\chi_\perp(\omega)$ has not been previously reported, and is the main contribution to $S(\omega)$.  The susceptibilities peak at different frequencies, $\omega_{\alpha\alpha}$ and $\omega_\perp$, whose temperature dependence and relation to spectral peaks are shown in Fig. \ref{fig:peaks}.  

At low temperatures, $\chi_{\alpha\alpha}''(\omega)$ are inelastic, with peaks at $\omega_{11} \approx 1.3 \omega_K$, $\omega_{22} \approx 1.3 \omega_\Delta$, where the proportionality factor for these peaks is consistently between $1.2-1.4$ at low temperatures; this proportionality is similar to isotropic $SU(N)$ Anderson results \cite{Bickers1987}.  These responses arise from magnetic excitations between the ground state Kondo singlet and the low-lying magnetic doublets.  These magnetic excitations occur at slightly enhanced frequencies compared to the charge excitations measured by $A_{\alpha\alpha}(\omega)$, but represent the same Kondo resonances\cite{Bickers1987}.  As $T$ increases, there is a crossover from inelastic to quasi-elastic behavior, diagnosed by the absence of peaks in $\chi''(\omega)/\omega$ \cite{Hoshino2009,SM}.  $\chi_{11}''$ becomes quasi-elastic around $T_K$, while $\chi_{22}''$ becomes quasi-elastic between $T_K$ and $\Delta$.  

The inter-doublet susceptibility, $\chi''_\perp(\omega)$ captures fluctuations out of the ground state Kondo singlet into the excited states that change the orbital character.  For large $\Delta$, these are primarily excitations into excited states with $f_2$ character, which can in principle be either the excited magnetic doublet at $\omega_{22} \propto \omega_\Delta$ or the excited Kondo singlet at $\omega_\perp \propto \omega_E$.  With smaller $\Delta$, there is also $f_2$ weight in the ground state, allowing excitations into the lower magnetic doublet at $\omega_{11} \propto \omega_K$, as well as a small amount of fluctuations into the $f_1$ part of the excited Kondo singlet at $\omega_\perp$.  Overall, we find the INS structure factor to have up to two peaks at $\omega_{11}$ and $\omega_\perp$ with distinct temperature dependences [Fig. \ref{fig:INS}(b)].  
The lower peak is clearly associated with $\omega_K$ and shares its $T$ dependence, while the upper peak is generally at $\omega_\perp < \omega_{22}$.  Indeed, $\omega_\perp$ has much more in common with $\omega_E$ than with $\omega_\Delta$. Both $\omega_\perp$ and $\omega_E$ decrease with increasing $T$; increase with increasing $\Delta$; and remain basically flat with increasing $|\epsilon_f|$.  Both peaks survive without significant broadening to much higher temperatures than both $\omega_{\alpha\alpha}$ and the Kondo resonance peaks in $A_{\alpha \alpha}$.   

Therefore, we interpret $\omega_\perp$ as the energy difference between the ground and excited state Kondo singlets, which is directly measured by $\omega_E = \Delta^*-\omega_{Ke}$ in $A_{11}(\omega < 0)$. Unlike the proportionality factors between $\omega_K, \omega_{11}$ and $\omega_\Delta, \omega_{22}$, the proportionality factor between $\omega_E$ and $\omega_\perp$ varies widely between $1-2$, with larger values for larger $\Delta/\omega_K$. Interestingly, the monotonic temperature renormalization of $\omega_\perp$ follows the anisotropic $SU(4)$ poor man's scaling expectations \cite{Eix2024}.  For the symmetric hybridization considered here, $\Gamma_\alpha  = \Gamma_0$, there are two relevant Kondo couplings, $J_{\alpha}$ associated with the two doublets, $\frac{\pi}{2D} J_\alpha(T = \infty) = \Gamma_0/(|\epsilon_f|+\Delta \delta_{\alpha 2})$.  The CEF splitting is initially enhanced by valence fluctuations as $\partial \Delta/\partial \log D \propto -\rho (J_1^2 - J_2^2)$, consistent with the $T$ dependence of $\omega_\perp \sim \omega_E$.  By contrast, $S\geq 1$ anisotropic $SU(2)$ Kondo models show anisotropy switching even for initially isotropic cases \cite{Kwasigroch2022}, suggesting unexpected differences between the spin-$S$ $SU(2)$ Kondo relevant for transition metals\cite{Zitko2008,Otte2008,Zitko2009b,Hock2013,Kwasigroch2022} and the anisotropic $SU(N)$ Kondo relevant for rare earths.  

We expect the qualitative results from our minimal impurity model to apply broadly to real materials based on Ce, Yb and Sm.  Real materials in sufficiently low symmetries to fully split the ground state $J$ multiplet will of course have more doublets, with more CEF splittings $\Delta_j$, as well as slightly different structure factors.  Each CEF splitting should correspond to both a CEF peak in the appropriate $A_{\alpha \alpha}$, as well as excited Kondo singlet peaks appearing in $A_{11}$ and $S(w)$.  The larger $N = 2J+1$ will likely lead to larger $\omega_K$'s and more temperature enhancement.  Cubic materials will have both quartets and doublets, and materials with a ground state doublet and an excited state quartet should exhibit a \emph{larger} difference between $\Delta^*$ and $\omega_\perp$ due to the enhanced excited state Kondo temperature.  The applicability to Sm (4f$^5$, $J = 5/2)$ materials that fluctuate to 4$f^6$ ($J = 0$) requires $\omega_K$ to be sufficiently low that the excited 4$f^6$ $J = 1$ multiplet can be neglected.  It is an interesting open question how the renormalization of the magnetic anisotropy will manifest in Tm$^{3+}$ or U$^{4+}$, which have magnetic excited valence states and may exhibit multi-channel Kondo physics.  Anisotropic mixed valent materials, like CeRhSn \cite{Kim2003,Sundermann2021,Bohm2024} and $\beta$-YbAl$_4$ \cite{Nakatsuji2008,Okawa2010} may be good candidates to examine the temperature variation in $\Delta^*$ in RIXS measurements, although care must be taken to distinguish Kondo renormalization from more conventional thermal changes in the $c/a$ ratio. It will be particularly interesting to contrast different spectroscopic measurements for cases where $\Delta \lesssim T_K$; for example, the larger CEF splittings measured by RIXS compared to INS in CeCu$_2$Si$_2$ are consistent with our results \cite{Amorese2016}. The hexagonal CeRh$_6$Ge$_4$ would be also a good candidate, as it has two low-lying doublets split by $\Delta_1 = 5.8$meV on the order of $T_K = 4.5$meV, with a higher doublet at $\Delta_2 = 22$meV \cite{Shu2021,Thomas2024}. 

$SU(4)$ Anderson impurity physics is also found in carbon nanotube based quantum dots, where SOC plays a similar role to the CEF splitting~\cite{Choi2005,Jarillo-Herrero2005,Laird2015,Teratani2020}, although these systems have much smaller $U$.  The more symmetric nature of the finite-$U$ $SU(4)$ Anderson model will change the peak structure significantly, and how the peaks will renormalize is an interesting direction for future work.

To summarize, our results show distinct behavior in the CEF scales measured by electronic responses like photoemission, RIXS and STM ($\Delta^*$), compared to the scales measured by spin responses like INS ($\omega_\perp$).  This difference arises because electronic measurements measure the splitting of the Kondo resonance, while spin spectroscopies measure the splitting of the ground and excited state Kondo singlets. This deviation is minimal when $T_K/\Delta$ is small, but is enhanced when the valence fluctuations renormalize the two splittings differently, as seen in Fig. \ref{fig:enhance}.  Moreover, when there are significant valence fluctuations, both quantities show an appreciable temperature dependence, with most variation in $\Delta^*$ occurring at high temperatures, where the valence is changing, but most variation in $\omega_\perp$ occurring near $T_K$, where the spin is being screened.

This work sheds light on how various CEF measurements connect to underlying theoretical models, and when different experimental techniques really measure different physical quantities, as for $\Delta \lesssim T_K$ \cite{Amorese2016}, or the same, for $\Delta \gg T_K$.  Note that we choose isotropic hybridizations, $\Gamma_\alpha = \Gamma_0$, which likely lead to minimal renormalizations of $\Delta^*$, $\omega_\perp$ compared to real materials with anisotropic hybridizations.  The enhancement can be much larger ($\Gamma_2 < \Gamma_1$) or become negative ($\Gamma_2 > \Gamma_1$), driving CEF inversions \cite{Hattori2010,Pourovskii2014,Rueff2015,Kwasigroch2022}, as expected for CeRh$_6$Ge$_4$ \cite{Thomas2024}.  How these inversions manifest in the spectroscopic and thermodynamic responses is an interesting question for future work.  


\begin{acknowledgments}
We acknowledge stimulating discussions with Joseph Eix, Filip Ronning, and Benjamin Ueland. This work was supported by the U.S. Department of Energy, Office of Science, Basic Energy Sciences, Materials Science and Engineering.  Ames National Laboratory is operated for the USDOE by Iowa State University under Contract No. DE-AC02-07CH11358
\end{acknowledgments}

\bibliographystyle{apsrev4-1}
\bibliography{NRG.bib}

\clearpage
\onecolumngrid
\begin{center}
{\bf Supplemental Material: ``Distinct effects of Kondo physics on crystal field splitting in electronic and spin spectroscopies''}
\end{center}
\vspace{.5cm}%

In this supplemental material, we discuss how well our numerical renormalization group (NRG) results satisfy the expected sum rules in Section I; discuss the estimated numerical error in the spectral function peak locations in section II; provide several example spectral functions and structure factors to show the temperature evolution of $\Delta^*$ and $\omega_\perp$ discussed in the main paper in section III; provide examples of the evolution of dynamical susceptibilities from inelastic to quasi-elastic as a function of temperature in section IV; and finally provide additional details of the derivation of the inelastic neutron scattering response in section V.


\section{Friedel sum rule and other checks}

We expect the Friedel sum rule to be satisfied at zero temperature, which is a relationship between $A_{\alpha\alpha}(0)$ for a single spin component and the phase shifts,
\begin{equation}
    A_{\alpha\alpha}(0) = \frac{1}{\pi \Gamma_0}\sin^2\left(\frac{\pi n_\alpha}{2}\right).
\end{equation}
Here, we use the $A_{\alpha\alpha}(0)$ of the fitted functions for the lowest temperatures calculated ($T_0 = .0001 \Gamma_0$), and find that the Friedel sum rule is satisfied within $\approx 1-5\%$ for both $\alpha$, with the larger errors for the small $T_K$ cases where the errors are likely due to finite temperature.  These are not improved by increasing $m$ or decreasing the $\Lambda$.

We can also check the $\Delta = 0$ results with the $n_0$ for the Bethe ansatz $J=3/2$ calculation \cite{Schlottmann1989}.  Bethe ansatz predicts $n_f(T) = .653, .824, .911$ for $\epsilon_f = -4\Gamma_0, -6\Gamma_0, -8\Gamma_0$ respectively, and we find at $T_0 = .0001 \Gamma_0$: $n_f(T_0) = .651, .826, .913$.  The agreement of $A(\omega = 0)$ with the Friedel sum rule is $98.7\%, 99.8\%, 99.0\%$, respectively.

Bethe Ansatz also allows us to compare $\chi(T = 0) = (g\mu_B)^2 j(j+1)/(3 k_B T_K)$, using the Bethe Ansatz $T_K$ for $\Delta = 0$.  In our calculation, for $\Delta = 0$, this quantity is $5/2\chi_{11}'(\omega = 0)$, which we can obtain by Kramers-Kronig transformation of $\chi''(\omega)$: $\chi'(0) = \int_{-\infty}^\infty d\omega \chi''(\omega)/\omega$.  
The $\Delta = 0$ Bethe ansatz $T_K$ are $T_K = .78\Gamma_0, .16\Gamma_0, .033\Gamma_0$ for $\Delta = 0$, $\epsilon_f = -4\Gamma_0, -6\Gamma_0, -8\Gamma_0$, leading to $\chi(0)\Gamma_0 = 1.60, 7.69, 37.0$, while we find $\chi(0)\Gamma_0 = 1.68, 7.62 , 36.5$.  

Finally, we can check the Korringa-Shiba relation for our $\Delta = 0$ results using only the spin susceptibility, where $\chi_{11}'(\omega = 0)^2/\lim_{\omega \rightarrow 0} \chi_{11}''(\omega)/\omega = \pi/4$.  The relationship holds to $97.3\%, 95.7\%, 95.2\%$ for $\epsilon_f = -4\Gamma_0, -6\Gamma_0, -8\Gamma_0$.

\section{Error estimates}

\begin{figure}[!ht]
    \centering
    \includegraphics[width=.6\textwidth]{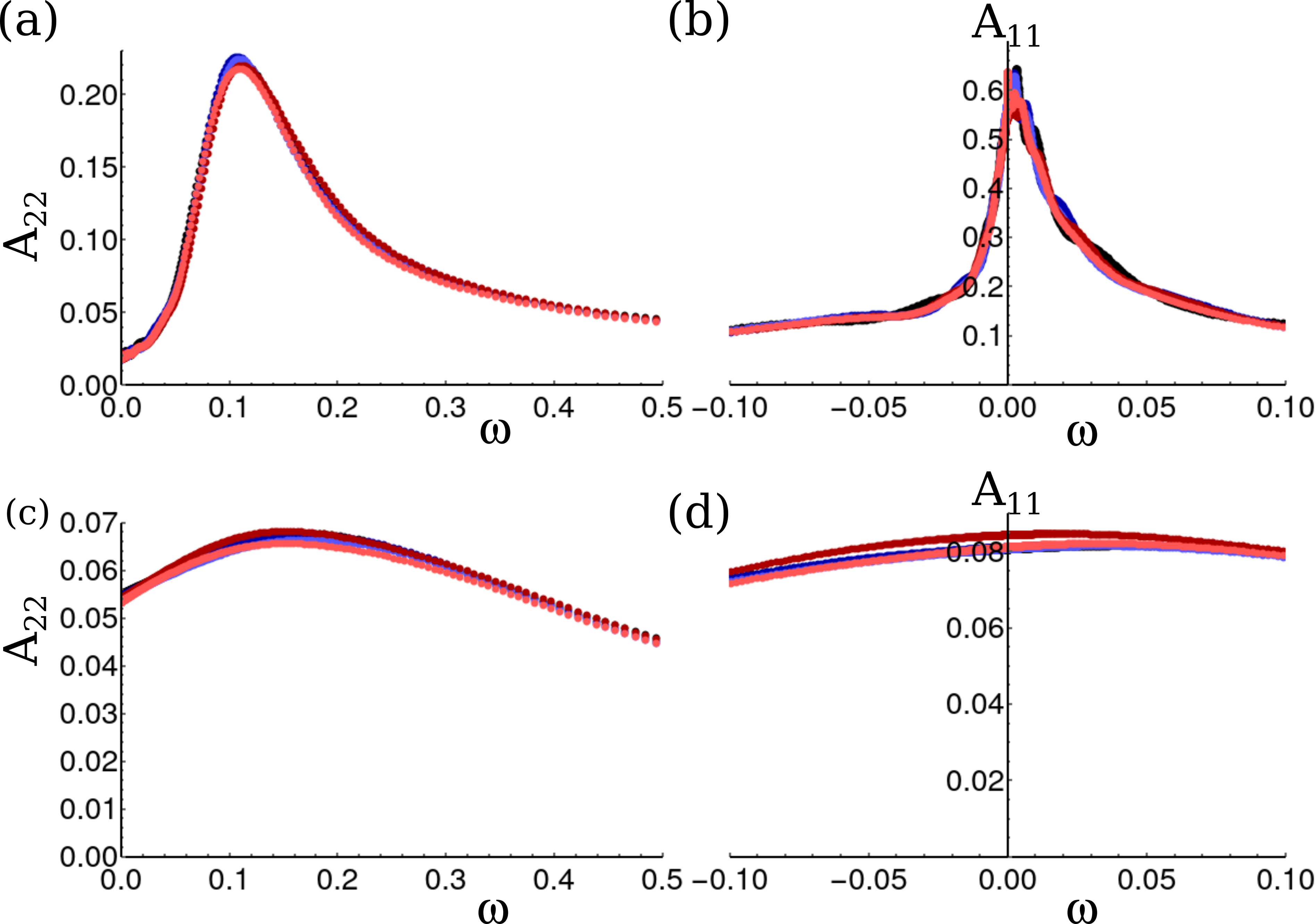}
    \caption{Error estimates for $\Delta = .1 \Gamma_0$, $\epsilon_f = -8\Gamma_0$. Five spectral function curves vs $\omega$ (in units of $\Gamma_0$) are shown in each sub-figure, calculated for $\Lambda = 4$, $m$ = 1500, as in the main text (black); $\Lambda = 3.5$, $m$ = 1500 (dark blue); $\Lambda = 3.5$, $m$ = 2000 (light blue); $\Lambda = 2.5$, $m$ = 1500 (dark red); $\Lambda = 2.5$, $m$ = 2000 (light red). The top row is $T=.0001\Gamma_0$: (a) shows the peak in $A_{22}$, with average peak location and standard deviation: $\omega_\Delta 
 = .110 \pm .001$; (b) shows the Kondo resonance peak in $A_{11}$, with $\omega_K = .0030\pm.0001$.  The bottom row shows $T = .17\Gamma_0$. (a) is the peak in $A_{22}$, with $\omega_\Delta = .163\pm.009$; (b) is the Kondo resonance peak in $A_{11}$, with $\omega_K = .0325\pm.008$.}
    \label{fig:errorsp18}
\end{figure}

Most of the computations in this paper were run with cutoff $\Lambda = 4$, $m$ = 1500 kept states and $N_z = 4$ discretization realizations.  At these values, the $f$-sum rule and Friedel sum rule are well satisfied, but there can be considerable numerical artifacts in the low temperature, low frequency behavior.  In order to check convergence of the peak locations and to make a very crude error estimate, we ran a series of calculations for $\epsilon_f = -8 \Gamma_0$, $\Delta = 0.1 \Gamma_0, 1.0 \Gamma_0$ and $T = 0.0001\Gamma_0$ and $0.17 \Gamma_0$.  In these, we chose $\Lambda = 2.5, 3.5, 4$, $m$ = 1500 or 2000, and $N_z = 4,8$ discretization realizations.  Increasing $N_z$ from $4$ to $8$ made no observable differences.

For $\Delta = 0.1\Gamma_0$, the results are shown in Fig. \ref{fig:errorsp18}, and there was no systematic change in either peak location at either temperature.  We fit the peak locations for each data set, where the Kondo resonance peaks are fit with generalized Fano-Frota functions, and then the maximum of that function is taken as the peak location, in order to avoid the effects of low frequency numerical artifacts. The crystal field (CEF) peak is always just the frequency that maximizes $A_{22}(\omega)$.  We estimate that the low temperature peaks are at $\omega_K/\Gamma_0  =0.0030\pm 0.0001$ and $\omega_\Delta/\Gamma_0 = 0.110 \pm 0.001$, giving $\Delta^*/\Gamma_0 = 0.107 \pm .001$.  We estimate that the higher temperature peaks are at $\omega_K/\Gamma_0 = 0.0325 \pm 0.008$, and $\omega_\Delta/\Gamma_0 = 0.163 \pm 0.009$, giving $\Delta^*/\Gamma_0 = 0.13 \pm 0.01$.

\begin{figure}[!ht]
    \centering
    \includegraphics[width=.6\textwidth]{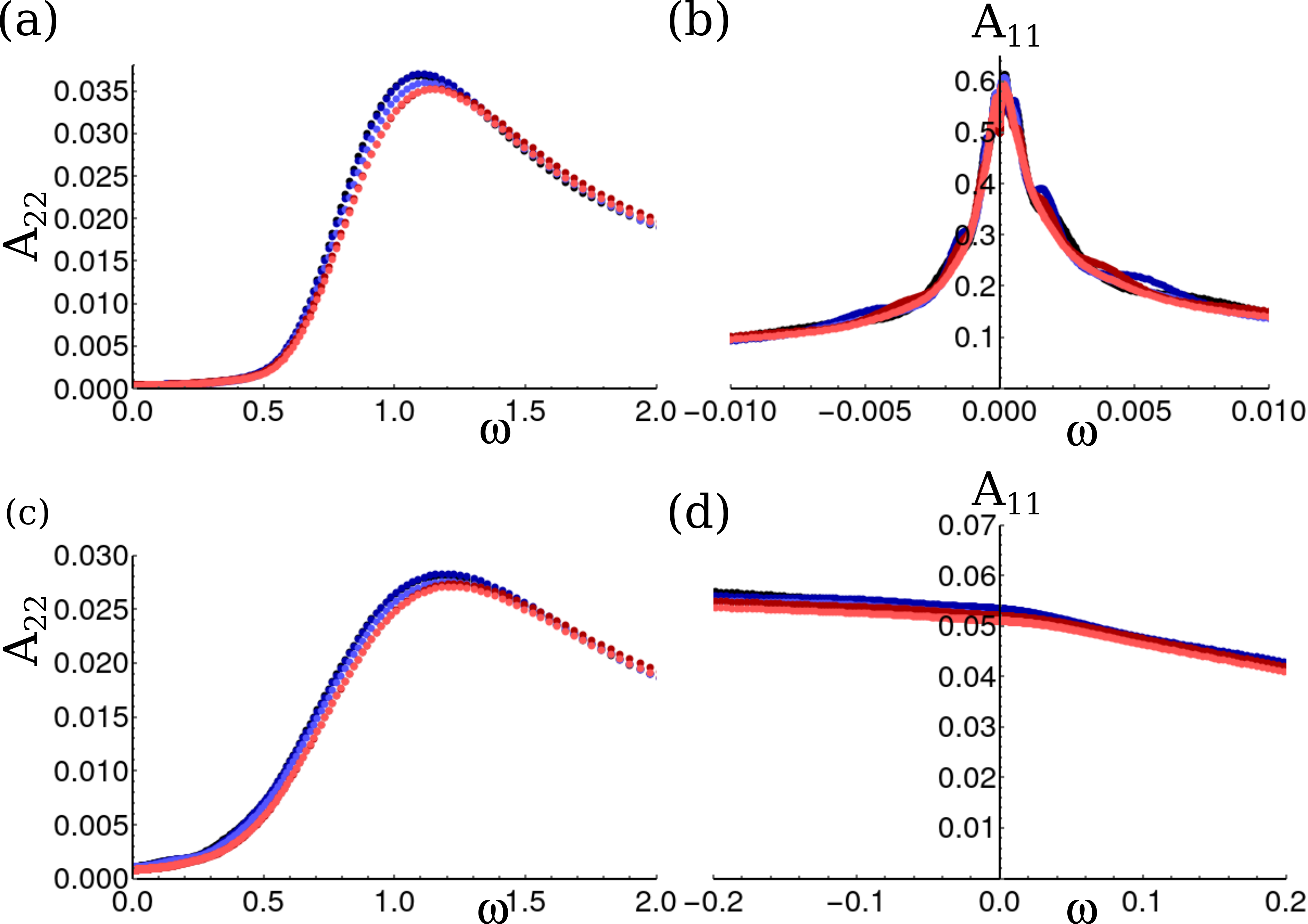}
    \caption{Error estimates for $\Delta = 1.0 \Gamma_0$, $\epsilon_f = -8\Gamma_0$. Five spectral function curves vs $\omega$ (in units of $\Gamma_0$) are shown in each sub-figure, calculated for $\Lambda = 4$, $m$ = 1500, as in the main text (black); $\Lambda = 3.5$, $m$ = 1500 (dark blue); $\Lambda = 3.5$, $m$ = 2000 (light blue); $\Lambda = 2.5$, $m$ = 1500 (dark red); $\Lambda = 2.5$, $m$ = 2000 (light red). The top row is $T=.0001\Gamma_0$: (a) shows the peak in $A_{22}$, with an average peak location and standard deviation: $\omega_\Delta = 1.13 \pm .02$; (b) shows the Kondo resonance peak in $A_{11}$, with $\omega_K = .00023\pm.00002$.  The bottom row shows $T = .17\Gamma_0$. (a) is the peak in $A_{22}$, with $\omega_\Delta = 1.21\pm.02$; (b) is the Kondo resonance ``peak'' in $A_{11}$, which is always at negative frequencies, and likely conflated with the excited Kondo resonance. The effective CEF splitting is just $\omega_\Delta$: $\Delta^*(.17\Gamma_0) = 1.21\pm.02$.}
    \label{fig:errors18}
\end{figure}

For $\Delta = 1.0\Gamma_0$, the results are shown in Fig. \ref{fig:errors18}.  Here, there is no systematic change in the Kondo resonance peaks, but the $\omega_\Delta$ peak location does systematically increase with decreasing $\Lambda$ for these $m$, suggesting that the number of kept states is insufficient for smaller $\Lambda$ values.  We did not find any significant change in the peak location as $m$  was increased from 1500 to 2000, but did find that $m$ = 3000 for $T=.0001\Gamma_0$, $\Lambda = 2.5$ moved the peak back to where it was for $\Lambda = 3.5$ with $m$ = 2000. The quantitative error estimate for $\omega_\Delta$ should be taken with a grain of salt, but qualitatively the errors are still quite small. 
We fit the peak locations for each data set and dids error analysis.  We estimate that the low temperature peaks are at $\omega_K/\Gamma_0  =0.00023\pm 0.0002$ and $\omega_\Delta/\Gamma_0 = 1.13 \pm 0.02$, giving $\Delta^*/\Gamma_0 = 1.13 \pm .02$.  At higher temperature, there is no Kondo resonance, and we find $\omega_\Delta/\Gamma_0 = 1.21 \pm 0.02$, giving $\Delta^*/\Gamma_0 = 1.21 \pm 0.02$ to be the enhanced CEF splitting, $\Delta^*$. 

We did not examine errors in the dynamical spin susceptibility peak locations as systematically, but they appear to be similar.

\section{Temperature evolution of the spectral function and structure factor}

\begin{figure}[!ht]
    \centering
    \includegraphics[width=.98\textwidth]{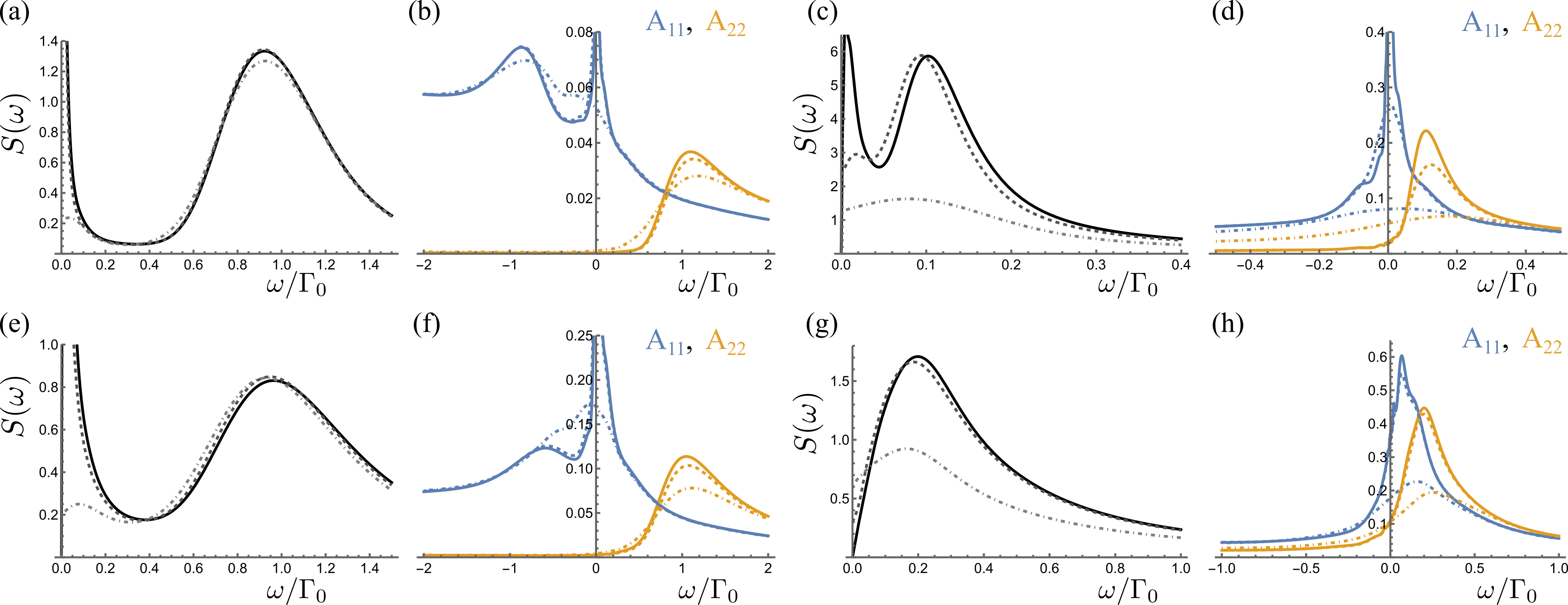}
    \caption{Temperature dependence of the spectral functions and structure factor for:  (a-b) $\Delta = \Gamma_0$, $\epsilon_f = -8\Gamma_0$; (c-d) $\Delta = .1\Gamma_0$, $\epsilon_f = -8 \Gamma_0$; (e-f) $\Delta = \Gamma_0$, $\epsilon_f = -6\Gamma_0$; and (g-h) $\Delta = .1\Gamma_0$, $\epsilon_f = -6\Gamma_0$.  For each pair, the left figure shows the structure factor, $S(\omega)$ used to calculate $\omega_\perp$ at three different temperatures  $T_0 = .0001\Gamma_0$ (black, solid); $T_8 = .02\Gamma_0$ (dark gray, dashed); and  $T_{11} = .17\Gamma_0$ (gray, dot-dashed), while the right figure shows the impurity spectral functions, $A_{\alpha\alpha}(\omega)$ used to calculate $\Delta^*$ at the same three temperatures ($T_0$ - solid, $T_8$ - dashed, $T_{11}$ - dot-dashed).  Compare to Figure \ref{fig:enhance} in the main text.}
    \label{fig:examples}
\end{figure}

In Fig. \ref{fig:examples}, we show the temperature evolution of the neutron scattering structure factor and impurity spectral functions for four cases: (a-b) $\Delta = \Gamma_0$, $\epsilon_f = -8\Gamma_0$; (c-d) $\Delta = .1\Gamma_0$, $\epsilon_f = -8 \Gamma_0$; (e-f) $\Delta = \Gamma_0$, $\epsilon_f = -6\Gamma_0$; and (g-h) $\Delta = .1\Gamma_0$, $\epsilon_f = -6\Gamma_0$.  The solid curves are the lowest temperature, $T_0 = .0001\Gamma_0$, which is always below the Kondo temperature; the dashed curves are $T_8 = .02\Gamma_0$, which is above the Kondo temperature except for (g-h); and the dot-dashed curves are for $T_{11} = .17\Gamma_0$, which is above all Kondo temperature scales and the bare $\Delta$ for cases (c-d) and (g-h).  

Except for (g-h), these examples all have $\Delta > \omega_K$ and exhibit $SU(2)$ Kondo physics, as can be seen from the distinction between the lower frequency Kondo resonance peaks and higher frequency CEF peaks seen in $S(\omega)$ at low temperatures.  For (g-h), there is only one broad peak in $S(\omega)$ with a peak at approximately $2\Delta$, while in $A(\omega)$, there is a clear splitting of the Kondo resonance by the CEF, with $\Delta^* \sim 1.2 \Delta$.

\begin{figure}[!ht]
    \centering
    \includegraphics[width=.8\textwidth]{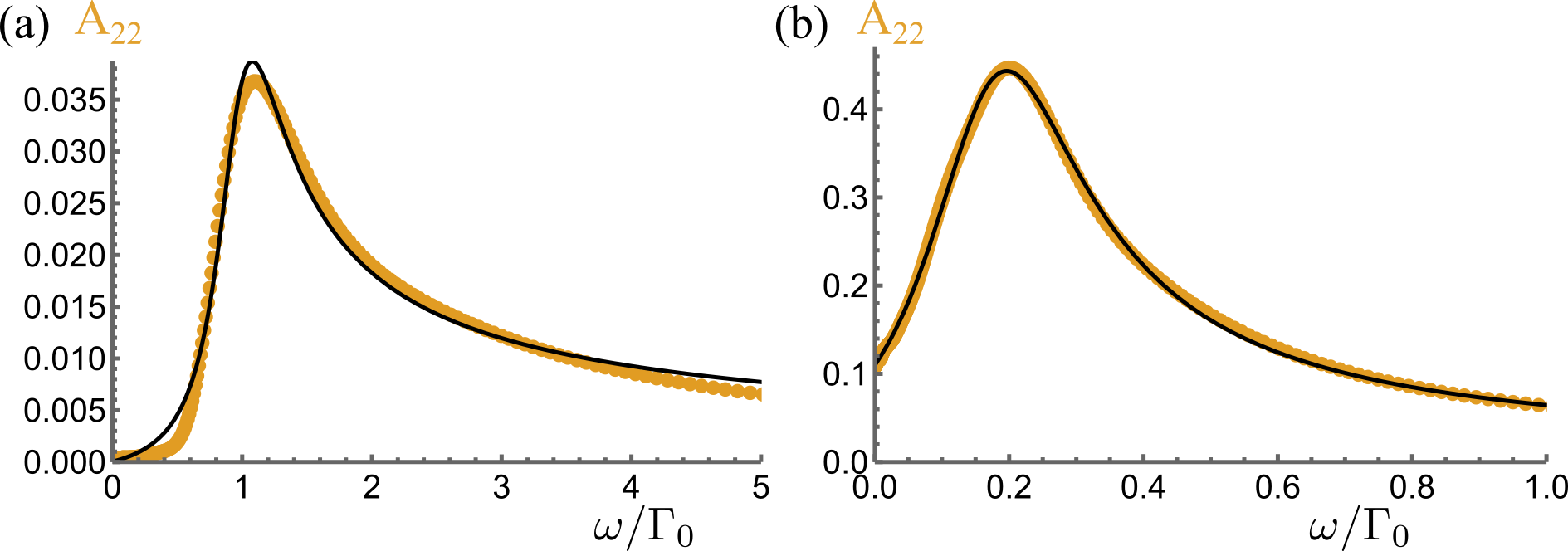}
    \caption{Generalized Fano-Frota fits, $A_{GF}(\omega) = A_0 \mathrm{Re}\left\{\mathrm{e}^{i\phi} \left[\frac{i \Gamma}{\omega-\omega_0 +i\Gamma}\right]^{a}\right\}$ for (a) $\Delta = \Gamma_0$, $\epsilon_f = -8\Gamma_0$ at $T_0 = .0001\Gamma_0$, where the fit is poor: fit parameters are: $A_0 = .046$, $\omega_0 = .96$, $\Gamma = .25$, $a = .64$, $\phi =.74$, all in units of $\Gamma_0$.  (b) $\Delta = 0.1\Gamma_0$, $\epsilon_f = -6\Gamma_0$ at $T_0 = .0001\Gamma_0$: fit parameters are: $A_0 =.46$, $\omega_0 = .16$, $\Gamma = .13$, $a = .88$, $\phi =-.43$.  The large distinction in weights between the two peaks is due to the very small value of $n_2 = .014$ ($n_1 = .92$) at low temperatures for case (a), versus $n_2 = .28$ ($n_1 = .55$) for case (b). }
    \label{fig:ff}
\end{figure}

The change in the shape of the peak in $A_{22}(\omega)$ between more $SU(2)$ and more $SU(4)$ physics is clear here.  In Fig. \ref{fig:ff}, we show the generalized Fano-Frota fits of $A_{22}(\omega)$ for case (a-b) and case (g-h), at $T_0$.  For case (a-b), the fit is poor and the data shows a clear gap at low frequencies, indicating this case does not behave like a Kondo resonance peak.  For case (g-h), the fit is quite good, indicating that this peak should be thought of as a crystal-field split Kondo resonance.  The generalized Fano-Frota fits are always good for the lower energy Kondo resonance.

\section{Inelastic versus quasi-elastic response}

\begin{figure}[!ht]
    \centering
    \includegraphics[width=.95\textwidth]{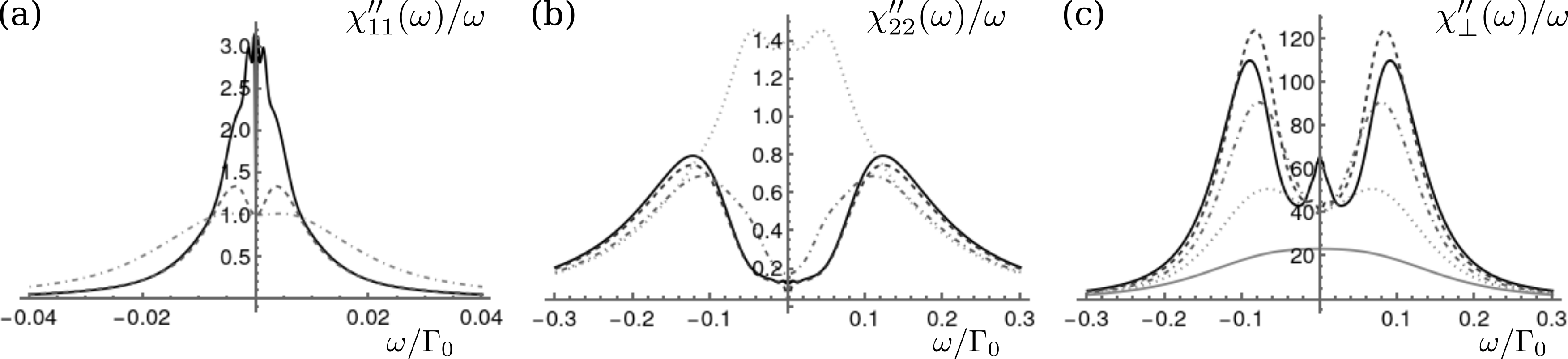}
    \caption{The distinct temperature dependences of the three dynamical spin susceptibilities,  $\chi''(\omega)/\omega$, for $\Delta = 0.1\Gamma_0$ and $\epsilon_f = -8 \Gamma_0$.  (a) $\chi''_{11}(\omega)/\omega$ normalized by its value at $\omega = 0$ shows how the signal for the Kondo resonance goes from inelastic to quasi-elastic as the temperature is raised from $T_0$ (solid, black) to $T_5$ (dark gray, dashed) to $T_7$ (gray, dot-dashed). (b) $\chi''_{22}(\omega)/\omega$ is dominated by the excited CEF scale, $\omega_\Delta$, and this behavior becomes quasi-elastic at a slightly higher temperature, well below $\Delta$, with the four curves shown at $T_0$ (solid, black), $T_5$ (dark gray, dashed) , $T_7$ (dark gray, dot-dashed), $T_8$ (gray, dotted) .  (c) $\chi''_\perp(\omega)/\omega$ has contributions from both the Kondo resonance and transition between the two Kondo resonances, $\omega_\perp$.  This peak remains quasi-elastic to much higher temperatures (roughly to $T = \Delta$) with the five curves shown at $T_0$ (solid, black), $T_7$ (dark gray, dashed) , $T_9$ (dark gray, dot-dashed), $T_{10}$ (gray, dotted), $T_{11}$ (gray).  The temperatures increase on a log scale, with $T_0 = .0001\Gamma_0$, $T_5 = .003\Gamma_0$, $T_7 = .01\Gamma_0$, $T_8 =.02\Gamma_0$, $T_9 = .04 \Gamma_0$, $T_{10} = .09 \Gamma_0$, $T_{11} = .17\Gamma_0$.  The Kondo temperature for this data set is roughly $T_5$, based on the temperature where $n_0(T)$ saturates.}
    \label{fig:qE}
\end{figure}

In order to understand the differences between the dynamical susceptibilities, it is useful to look at $\chi''(\omega)/\omega$, which will show a peak at finite frequencies for an inelastic response, but will have a peak at $\omega = 0$ for a Lorentzian quasi-elastic response.  As can be seen in Fig. \ref{fig:qE}, the three different susceptibilities all lose their inelastic nature at different temperatures, with $\chi_{11}$ becoming quasi-elastic at $\sim 10 T_K$, $\chi_{22}$ becoming quasi-elastic somewhere between $T_K$ and $\Delta$ and $\chi_{\perp}$ becoming quasi-elastic around $\Delta$.

\section{Inelastic neutron scattering}

Inelastic neutron scattering to detect CEF splittings typically measures the powder average of,
\begin{align}
    S(\omega, \vect{Q}) & \propto n_B(\omega) \chi''(\omega,\vect{Q}) \cr
    & \propto g_L^2 \sum_{a,b}n_a |\langle a|\vect{J}_\perp|b\rangle|^2 \delta(E_b-E_a-\hbar \omega)
\end{align}
Here, $n_B$ is the Bose function, $g_L$ is the Lande g-factor, $n_a$ is the thermal occupancy of $|a\rangle$, and $a,b$ will cover the $J_z = \pm 1/2, \pm 3/2$ CEF states.  $\vect{J}_\perp = \hat Q \times (\vect{J}\times \hat Q)$ is the perpendicular component of $\vect{J}$.  The powder average in a tetragonal sample gives,
\begin{equation}
    \chi''(\omega) \propto \sum_{a,b}n_a \left[\frac{2}{3}|\langle a|J_x|b\rangle|^2+\frac{1}{3}|\langle a|J_z|b\rangle|^2\right] \delta(E_b-E_a-\hbar \omega).
\end{equation}
We can obtain this structure factor in our calculations, for a ground state doublet $|\pm 3/2\rangle$ and excited state doublet $|\pm 1/2\rangle$ as,
\begin{align}
    S(\omega) n_B^{-1}(\omega) \propto \frac{1}{3}\left[\left(\frac{3}{2}\right)^2\chi''_{S,11}(\omega)+\left(\frac{1}{2}\right)^2\chi''_{S,22}(\omega)\right] +\frac{2}{3}\left[\chi''_{S,22}(\omega) + \left(\frac{\sqrt{3}}{2}\right)^2\chi''_\perp(\omega) \right].
\end{align}
Here, $\chi_{S,\alpha\alpha}(\omega)$ is the $S_z$ spin-susceptibility and $\chi_\perp(\omega)$ is a cross-doublet contribution to the $S_x$ spin-susceptibility, where $J_x = \sum_{\sigma} f\dg_{1\sigma}f_{2\sigma} + f\dg_{2\sigma}f_{1\sigma}$ is used to calculate $\chi_\perp(\omega) = \int dt \langle J_x(t)J_x(0)\rangle \mathrm{e}^{-i \omega t}$.  We calculate all of these quantities within NRG to obtain the impurity structure factor $S(\omega)$ at different temperatures.  Note that the form of this expression only depends on the identity of the ``ground'' and ``excited'' state doublets, not their actual ordering, which may evolve with temperature.


\end{document}